\documentclass[11pt]{article}%
\usepackage{amssymb,amsmath,mathtools,xcolor,graphicx,xspace,colortbl,ragged2e,rotating}
\usepackage[margin=2.5cm]{geometry}
\usepackage{amsfonts}
\usepackage{amsmath}
\usepackage{amssymb}
\usepackage{graphics}
\usepackage{graphicx}%
\providecommand{\U}[1]{\protect\rule{.1in}{.1in}}
\graphicspath{{Manuscript9_graphics/}{Manuscript9_tcache/}{Manuscript9_gcache/}}
\DeclareGraphicsExtensions{.pdf,.eps,.ps,.png,.jpg,.jpeg}
\providecommand{\U}[1]{\protect\rule{.1in}{.1in}}

\begin{document}

\title{The Universe as a Learning System}
\author{Tomer Shushi\\Center for Quantum Science and Technology\\\& Department of Business Administration,\\Guilford Glazer Faculty of Business and Management,\\Ben-Gurion University of the Negev, Beer-Sheva, Israel}
\maketitle

\begin{abstract}
At its microscopic level, the universe follows the laws of quantum mechanics.
Focusing on the quantum trajectories of particles as followed from the
hydrodynamical formulation of quantum mechanics, we propose that under general
requirements, quantum systems follow a disrupted version of the gradient
descent model, a basic machine learning algorithm, where the learning is
distorted due to the self-organizing process of the quantum system. Such a
learning process is possible only when we assume dissipation, i.e., that the
quantum system is open. The friction parameter determines the nonlinearity of
the quantum system. We then provide an empirical demonstration of the proposed model.
\end{abstract}

The universe at its microscopic scales is modeled by quantum mechanics, which
provides a complete description of the particles that are the building blocks
of the universe. One of the main open issues in quantum mechanics is the
inability to determine the physical nature of quantum particles. In the
Schr\"{o}dinger picture, the quantum particles are described by wavefunctions,
followed by the Schr\"{o}dinger equation. However, the onticity of the
wavefunction remains a topic of ongoing debate. Another less-explored
formulation of quantum mechanics is the hydrodynamical formulation, which was
introduced by Erwin Madelung a year after Erwin Schr\"{o}dinger published his
famous equation. In the Madelung formulation, the quantum particles are
described by a (quantum) fluid [1-4].

Suppose we have a quantum system of $N$ one-dimensional non-relativistic
particles with positions $x_{1},x_{2},...,x_{N}$ and equal mass of $m.$ Then,
the Schr\"{o}dinger equation takes the form $i\hbar\partial_{t}\psi
=-\frac{\hbar^{2}}{2m}\nabla^{2}\psi+V\psi$ where $\psi=\psi\left(
\boldsymbol{x},t\right)  ,$ $\boldsymbol{x}=\left(  x_{1},x_{2},...,x_{N}%
\right)  ,$ is the wavefunction of the system, and $\nabla^{2}=-\sum_{j=1}%
^{N}\frac{\partial^{2}}{\partial x_{j}^{2}}.$ We assume that $\psi$ is a
smooth function\ and consider the polar representation
\begin{equation}
\psi\left(  \boldsymbol{x},t\right)  =R\left(  \boldsymbol{x},t\right)
e^{iS\left(  \boldsymbol{x},t\right)  /\hbar}, \label{01}%
\end{equation}
where $\rho=R^{2}$ is the probability density function (pdf) of the quantum
particle, $S$ is the phase, and we define the flow velocity by
$\boldsymbol{u=}\nabla S/m$. Then, by substituting (\ref{01}) into the
Schr\"{o}dinger equation, we obtain the hydrodynamical formulation of the
quantum particles which are given by the continuity equation%
\begin{equation}
\partial_{t}\rho+\nabla\cdot\left(  \rho\boldsymbol{u}\right)  =0, \label{02}%
\end{equation}
and the quantum Hamilton-Jacobi equation%
\begin{equation}
\partial_{t}\boldsymbol{u}+\boldsymbol{u}\cdot\nabla\boldsymbol{u}=-\frac
{1}{m}\nabla\left(  Q+V\right)  . \label{03}%
\end{equation}
Eq. (\ref{02}) promises the conservation of the pdf $\rho$ over time, and
(\ref{03}) is the classical Hamilton-Jacobi equation of the fluid, but, with
the addition of the quantum potential%
\begin{equation}
Q=-{\frac{\hbar^{2}}{2m}}{\frac{\nabla^{2}R\left(  \boldsymbol{x},t\right)
}{R\left(  \boldsymbol{x},t\right)  }.} \label{04}%
\end{equation}
The quantum potential (\ref{04}) can be seen as the self-organized process of
the quantum system, as it\ provides\ a universal connection between the
density function $\rho$ and the flow velocity $\boldsymbol{u}$ of the quantum fluid.

Self-organizing processes refer to systems that spontaneously arrange
themselves without external direction or control [5-6]. This phenomenon is
often observed in various natural and artificial systems, ranging from
biological organisms to social networks and computer algorithms.
Self-organization arises from simple interactions and feedback mechanisms
among individual components, leading to emergent patterns or structures at a
higher level of complexity. In the following, we examine how quantum mechanics
can be described as a learning system followed by quantum trajectories derived
from the hydrodynamical formulation of the quantum particles.

The connection between self-organizing processes and learning is fundamental
in artificial intelligence [7-9]. Self-organizing processes, characterized by
autonomous organization driven by internal dynamics, are integral to learning
paradigms such as unsupervised and semi-supervised learning. AI systems can
autonomously adapt and improve over time by integrating self-organizing
mechanisms into learning algorithms, enhancing adaptability and performance
across diverse domains. Learning systems often exhibit self-organizing
properties as they iteratively process data, make predictions, receive
feedback, and adjust their internal parameters accordingly.

The gradient descent algorithm is a foundational optimization technique widely
employed in machine learning to minimize the loss function associated with
training a model. It operates iteratively, adjusting the model's parameters in
a manner opposite to the gradient of the loss function, progressively moving
towards the minimum of the function. The standard gradient descent follows the
formula $x_{t+1}=x_{t}-\alpha\nabla f\left(  x_{t}\right)  ,$ $\alpha>0,$
$t=1,2,...,$ for minimizing the convex function $f$. The momentum gradient
descent algorithm introduces a momentum term to improve the optimization
process (see, e.g., [10]). This momentum term accumulates the gradients of
previous iterations, effectively imparting inertia to the optimization
procedure. By incorporating information from past gradients, momentum gradient
descent achieves smoother and more consistent updates to the model parameters,
accelerating convergence, especially in regions with high curvature or sparse gradients.

The mechanism of momentum gradient descent can be conceptualized as a
classical system of a ball rolling down a hill. As the ball gathers momentum
from its past movement, it continues in the same direction, even if the
gradient momentarily suggests otherwise.

The momentum gradient descent algorithm is given by%
\begin{equation}%
%TCIMACRO{\QATOPD{\{}{.}{x_{t+1}=x_{t}+v_{t}}{v_{t}=\beta v_{t-1}-\alpha\nabla
%f\left(  x_{t}\right)  }}%
%BeginExpansion
\genfrac{\{}{.}{0pt}{}{x_{t+1}=x_{t}+v_{t}}{v_{t}=\beta v_{t-1}-\alpha\nabla
f\left(  x_{t}\right)  }%
%EndExpansion
,\text{ }t=1,2,..., \label{GDm1}%
\end{equation}
for an initial datum $x_{0},v_{0}.$

The model can be derived by assuming that we have a classical particle that is
in a potential field $f\left(  x_{t}\right)  ,$ and so the force acting on the
particle is $F=-\nabla f\left(  x_{t}\right)  .$ Then, following Newton's law
of motion,%
\[
F=ma,
\]
where $a$ is the acceleration of the particle, by integrating Newton's
equation for a small unit of time, we obtain the equation for the velocity%
\begin{equation}
v_{t}=v_{t-1}-\frac{1}{m}\nabla f\left(  x_{t}\right)  -\mu v_{t-1}
\label{vt1}%
\end{equation}
where we have added $-\mu v_{t-1},\mu>0,$ to consider a friction term. The
integration of (\ref{vt1}) leads to the particle's position $x_{t+1}%
=x_{t}+v_{t},$ which gives the first equation in the momentum gradient
descent\ model, where the second equation is (\ref{vt1}) with $\beta=1-\mu$
and $\alpha=1/m.$

Going back to quantum mechanics, we can define the trajectory momentum through
the flow velocity $p\left(  \boldsymbol{x},t\right)  =m\cdot\boldsymbol{u}$,
and substitute it into (\ref{03}), we obtain the equation of motion for the
trajectory momentum
\begin{equation}
\partial_{t}\boldsymbol{p}+\frac{1}{m}\boldsymbol{p\cdot}\nabla\boldsymbol{p}%
=-\nabla\left(  Q+V\right)  .\label{Eq_Madp1}%
\end{equation}
We can then set the quantum trajectory $\left(  \boldsymbol{x}_{t}%
,\boldsymbol{p}_{t}\right)  $ that follows Newton's equation of motion, which
yields to $\frac{d}{dt}\boldsymbol{p}_{t}=-\nabla\left(  Q+V\right)
|_{\boldsymbol{x}=\boldsymbol{x}_{t}}$ where $\frac{d}{dt}\boldsymbol{x}%
_{t}=\frac{1}{m}\boldsymbol{p}_{t}$ (see, [11]).$\ $We now consider\ the
discretization of time for a small unit of time $\Delta t=1,$ and similar to
the case of a classical particle with friction, we add a friction term
$-\mu\boldsymbol{p}_{t-1}$ to eq. (\ref{Eq_Madp1}).\ Considering the flow
velocity $\boldsymbol{u,}$ we then obtain the equation%
\begin{align}
\boldsymbol{u}_{t} &  =\boldsymbol{u}_{t-1}-\frac{1}{m}\nabla\left(
Q+V\right)  -\mu\boldsymbol{u}_{t-1}\label{u_t1}\\
&  =\beta\boldsymbol{u}_{t-1}-\frac{1}{m}\nabla\left(  Q+V\right)  .\nonumber
\end{align}
Following the same discretization for the trajectory, we obtain the learning
equations%
\begin{equation}%
%TCIMACRO{\QATOPD{\{}{.}{\QTR{bs}{x}_{t+1}=\QTR{bs}{x}_{t}+\QTR{bs}{u}_{t}%
%}{\QTR{bs}{u}_{t}=\beta\QTR{bs}{u}_{t-1}-\lambda\nabla V+Dis_{t}\left(
%\QTR{bs}{x}_{t}\right)  }}%
%BeginExpansion
\genfrac{\{}{.}{0pt}{}{\boldsymbol{x}_{t+1}=\boldsymbol{x}_{t}+\boldsymbol{u}%
_{t}}{\boldsymbol{u}_{t}=\beta\boldsymbol{u}_{t-1}-\lambda\nabla
V+Dis_{t}\left(  \boldsymbol{x}_{t}\right)  }%
%EndExpansion
,\label{SQGD1}%
\end{equation}
with the learning parameter%
\begin{equation}
\lambda=\frac{1}{m},\label{Lambda1}%
\end{equation}
and a quantum term that disrupts the learning, $Dis_{t}\left(  \boldsymbol{x}%
_{t}\right)  ,$ that is given by
\begin{equation}
Dis_{t}\left(  \boldsymbol{x}_{t}\right)  =\frac{\hbar^{2}}{2m^{2}}\nabla
\frac{\nabla^{2}R\left(  \boldsymbol{x},t\right)  }{R\left(  \boldsymbol{x}%
,t\right)  }|_{\boldsymbol{x}=\boldsymbol{x}_{t}}.\label{D1}%
\end{equation}
The set of equations (\ref{SQGD1}) obtains the gradient descent learning
system for the goal function $V$ with the learning parameter $\lambda$. Unlike the classical case (\ref{GDm1}), here the learning is disrupted by a
quantum component $Dis_{t}\left(  \boldsymbol{x}_{t}\right)  ,$ that distorts
the learning process. 

Classicality can be achieved when taking $\hbar
\rightarrow0,$ which, in that case, the quantum disruptor vanishes,
\begin{equation}
\lim_{\hbar\rightarrow0}Dis_{t}\left(  \boldsymbol{x}_{t}\right)  =0.
\label{limD1}%
\end{equation}
We note that $Dis_{t}\left(  \boldsymbol{x}_{t}\right)  $ also decreases in
the case of larger mass $m,$ but it also implies a slower learning rate since
$\lambda$ is getting smaller as $m$ increases.

Adding a friction term to the classical description of the quantum particle is
necessary to achieve a learning mechanism. Similar to the derivation of the
momentum gradient descent, the particle will oscillate in case of no friction,
and the learning mechanism will break down. Such an addition may be seen as
artificial. However, there is a clear physical meaning for the quantum system.
It implies that the quantum system is an open system that dissipates, where
the modified (nonlinear) Schr\"{o}dinger equation is then given by
$i\hbar\partial_{t}\psi=-\frac{\hbar^{2}}{2m}\nabla^{2}\psi+V\psi+\mu\left(
S-\left\langle S\right\rangle \right)  \psi,$ and the value of the friction
parameter $\mu$ dictates the nonlinearity of the Schr\"{o}dinger equation
(see, again, [11]).

Let us now examine the proposed\ learning mechanism (\ref{SQGD1}) for a
specific\ quantum system. Suppose we have a single one-dimensional quantum
particle with mass $m=1$ that is subjected to the external potential $V\left(
x\right)  =\frac{1}{2}\omega^{2}x^{2}\ $and a friction $\mu=1.$ A solution of
the nonlinear Schr\"{o}dinger equation takes the form $\psi\left(  x,t\right)
=\sqrt[4]{\frac{\omega}{\pi}}e^{-\frac{\omega}{2}\left(  x-x_{t}\right)
^{2}+ip_{t}\left(  x-x_{t}\right)  +is_{t}}$ for $ds_{t}/dt=p_{t}^{2}%
/2-\omega^{2}x_{t}^{2}/2-\omega/2$ [11].

Then, the quantum disruptor is algebraically vanished,
\begin{equation}
Dis_{t}\left(  \boldsymbol{x}_{t}\right)  \equiv0, \label{DExample1}%
\end{equation}
obtaining a learning system without any disruptions.

In the following, we illustrate the convergence of the quantum trajectory
$x_{t}$ to the minimum point of the function $V$.

{\includegraphics[
height=2.2916in,
width=5.6382in
]{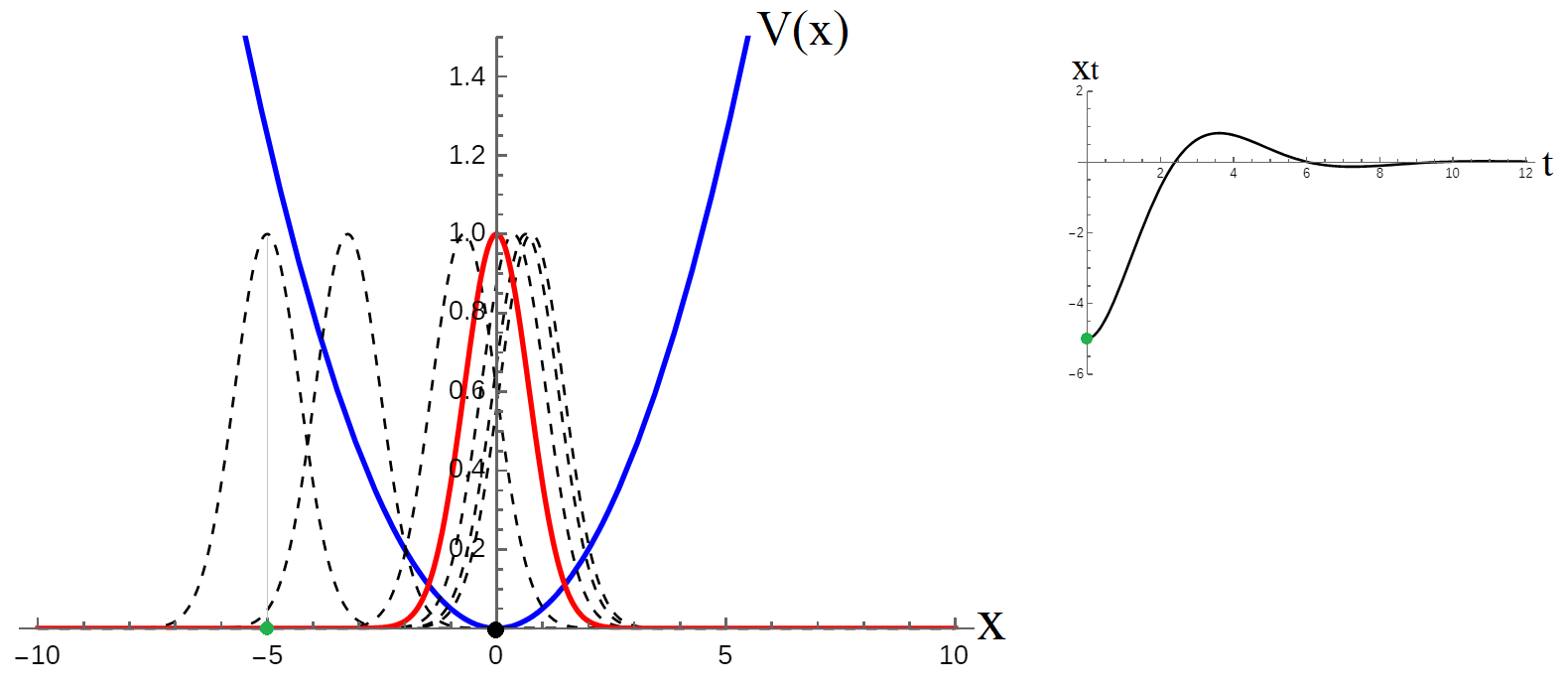}}

Figure 1. Illustration of learning for a quantum particle. Starting from the
Gaussian density function $\rho$ located around\ $x_{0}=-5,$ the location of
the density function converges to $x=0,$ which is the minimum point of the
potential $V=\frac{1}{2}\omega^{2}x^{2}.$ The small Figure illustrates the
dynamics of $x_{t}$ as it starts from $x_{0}=-5$ and converges to zero.

In conclusion, we have shown that quantum mechanics can be described as a
learning process once we adopt the Madelung formalism. Such a learning process
is possible only when we modify the quantum system by assuming dissipation,
i.e., that the quantum system is open. We conjecture that learning and
self-organizing processes have deep relations in the context of quantum
mechanics, followed by the quantum potential that disrupts learning. We have
demonstrated such a learning process through a gradient descent algorithm. We
note that a different theory has been proposed for cosmology in which the
universe learns its physical laws [12]. We further note that in quantum
computing, quantum gradient descent is a well-studied algorithm, and it
differs from the proposed learning system, which is focused on the universe as
a learning mechanism and not on artificial set-ups that involve classical and
quantum computers to exploit quantum effects for the sake of computing. We
hope that the proposed model will open the door for exploring more advanced
machine learning algorithms concealed in open quantum systems.

\end{document}